\documentclass[epsfig,color]{elsart}
\usepackage{graphicx}

\begin{document}
\begin{frontmatter}
\title{Upper limit on the neutrino magnetic moment from three years of
    data from the GEMMA spectrometer}

\author[ITEP]{A. G. Beda},
\author[JINR]{V. B. Brudanin},
\author[JINR]{V. G. Egorov\thanksref{email}},
\author[JINR]{D. V. Medvedev},
\author[YPI]{V. S. Pogosov},
\author[JINR]{M. V. Shirchenko},
\author[ITEP]{A. S. Starostin}
\thanks[email]{Corresponding author; {\em e-mail:} egorov@nusun.jinr.ru}
\address[ITEP] {State Scientific Center, Institute for
 Theoretical and Experimental Physics, Moscow, Russia}
\address[JINR] {Joint Institute for Nuclear Research, Dubna, Russia}
\address[YPI] {Yerevan Physics Institute, Yerevan, Armenia}
\begin{abstract}
 The result of the 3-year neutrino magnetic moment
measurement at the Kalinin Nuclear Power Plant ({\sl KNPP})
with the {\sl GEMMA} spectrometer is presented. Antineutri\-no-electron scattering is investigated. A high-purity
germanium detector of 1.5 kg placed at a distance of 13.9 m from the centre of the
3~GW$_{\rm th}$ reactor core is used in the spectrometer. The antineutrino
flux is 2.7$\times$10$^{\rm 13}\; \bar\nu_e$/cm$^2$/s. The
differential method is used to extract $\nu${\em -e}
electromagnetic scattering events. The scattered electron spectra
taken in 5184+6798 and 1853+1021 hours during the reactor {\sl ON} and {\sl
OFF} periods respectively are compared. The upper limits for the neutrino
magnetic moment $\mu _{\nu}$ with and without atomic ionization mechanism were found to be 5.0$\times$10$^{-12}\mu_{\rm B}$ and 3.2$\times$10$^{\rm -11}\mu_{\rm B}$ at 90{\%} CL, respectively.
\end{abstract}

\end{frontmatter}
\section{Introduction}

The Minimally Extended Standard Model predicts a very small magnetic moment for the massive neutrino ($\mu _\nu\sim10^{-20}\mu _{\rm B}$) which cannot be observed in any experiment at present.
On the other hand, there is a number of extensions of
the theory beyond the Minimal Standard Model where the {\sl Majorana} neutrino magnetic moment (NMM) could be at a level of
$10^{-(10\ldots12)}$~$\mu_{\rm B}$
irrespectively of the neutrino mass \cite{Voloshin86,Fukugita87,Pakvasa03,Gorchtein06,Bell06,Studenikin08,Giunti08,Giunti10}.
At the same time, from general considerations \cite{Bell05,Bell06d} it follows that
the {\sl Dirac} NMM cannot exceed $10^{-14}\mu_{\rm B}$.
Therefore, observation of the NMM value higher than $10^{-14}$~$\mu_{\rm B}$ would be evidence for New Physics and, in addition, would indicate \cite{Kayser08,Studenikin08,Giunti08,Giunti10} doubtless that the neutrino is a Majorana particle.

It is rather important to make laboratory NMM measurements sensitive enough to reach the $\sim$10$^{-11}\mu_{\rm B}$ region. The Savanna River experiment by Reines' group could be considered the beginning of such measurements. Over a period of thirty years the sensitivity of reactor experiments only increased by a factor of three: from $(2\ldots4)\times10^{-10}\mu_{\rm B}$\cite{Reines76,Vogel89} to $(6\ldots7)\times10^{-11}\mu_{\rm B}$\cite{TEXONO06,GEMMA07}. Similar limits were obtained for solar neutrinos\cite{SK04,BOREXINO}, but due to oscillations at long distance (as well as matter-en\-han\-ced oscillations in the Sun)
their flavor composition changes and therefore the solar NMM results
could differ from the reactor ones.

In this paper, the results of a 3-year NMM measurement being performed by the collaboration of
ITEP (Moscow) and JINR (Du\-b\-na) are presented.
Measurements are carried out with the {\sl GEMMA} spectro\-meter\cite{Beda98,Beda04,GEMMA07} at the 3~GW$_{\rm th}$ reactor of
the Kalinin Nuclear Power Plant (KNPP).

\section{Experimental approach}
A laboratory measurement of the NMM is based on its contribution to the $\nu$-$e$
scattering. For nonzero NMM the $\nu$-$e$ differential cross
section is given \cite{Vogel89} by a sum of the {\sl weak} interaction cross section
($d\sigma_W/dT$) and the {\sl electromagnetic} cross section ($d\sigma_{EM}/dT$):
\begin{figure}[hb]
 \setlength{\unitlength}{1mm}
  \begin{picture}(165,52)(0,0)
   \put(0,-2){\includegraphics{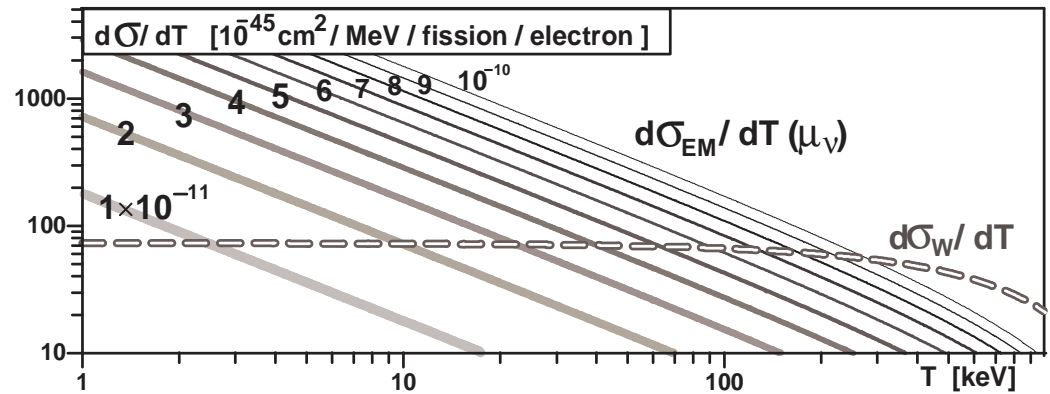}}
  \end{picture}
 \caption{Weak (W) and electromagnetic (EM) cross sections calculated for several
 NMM values.}
 \label{Fig.W/EM}
\end{figure}

\begin{equation}
 \label{eq.dsW/dT}
 \begin{array}{lll}
 {d\sigma_W}/{dT} &= \frac{G_F^2 m_e }{2\pi}&\left[ \left(1\!-
 \frac{T}{E_\nu}\right)^{\!2}\left(1\!+2\sin^2\theta_W\right)^{\!2} \!\!+ \right. \\
 &&\left.+\; 4\sin^4\theta_W -\! 2\left(1\!\!+2\sin^2\theta_W\right)\sin^2\theta_W
 \;\frac{m_eT}{E_\nu^2}\right],\\
 \end{array}
 \end{equation}
\begin{equation}
{d\sigma_{EM}}/{dT} =\pi r_0^2 \left( {\frac{\mu _\nu }{\mu _{\rm B} }}
\right)^2\left( {\frac{1}{T} - \frac{1}{E_\nu }}
\right)\;\;,
\label{eq.dsEM/dT}
\end{equation}

where $E_\nu$ is the incident neutrino energy, $T$ is the electron recoil
energy, $\theta_W$ is the Weinberg angle and $r_0$ is the electron radius ($\pi r_0^2=2.495\times10^{-25}$ cm$^2$).

Figure~\ref{Fig.W/EM} shows differential cross sections (\ref{eq.dsW/dT})
and (\ref{eq.dsEM/dT}) averaged over the typical antineutrino reactor spectrum
vs the electron recoil energy.
One can see that at a low recoil energy ($T\ll E_\nu$) the value of $d\sigma_W/dT$ becomes
almost constant, while $d\sigma_{EM}/dT$ behaves as $T^{-1}$, so that the
lowering of the detector threshold leads to a considerable increase in
the NMM effect with respect to the weak unremovable contribution.

To realize this useful feature in our GEMMA spec\-tro\-meter\cite{GEMMA07}, we use a 1.5 kg HPGe detector with an energy threshold as low as 3 keV. To be sure that there is no efficiency cut at this energy, the "hard" trigger threshold is made twice lower (1.5 keV).

\begin{figure}[ht]
 \setlength{\unitlength}{1mm}
  \begin{picture}(140,80)(0,0) 
   \put(5,2){\includegraphics{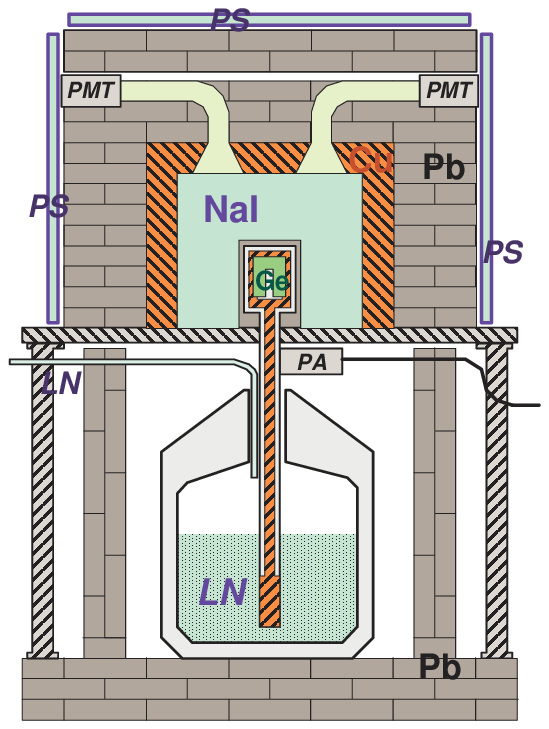}}
   \put(85,0){\includegraphics{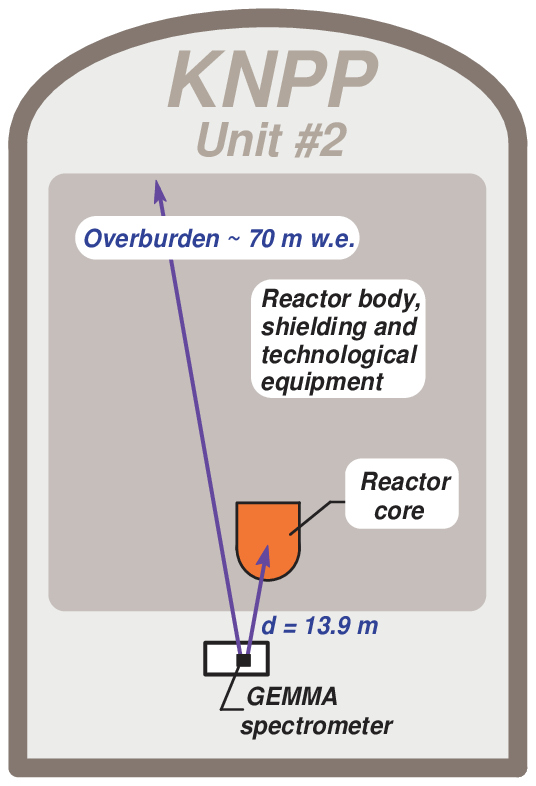}}
   \put(0,-1.5){\includegraphics{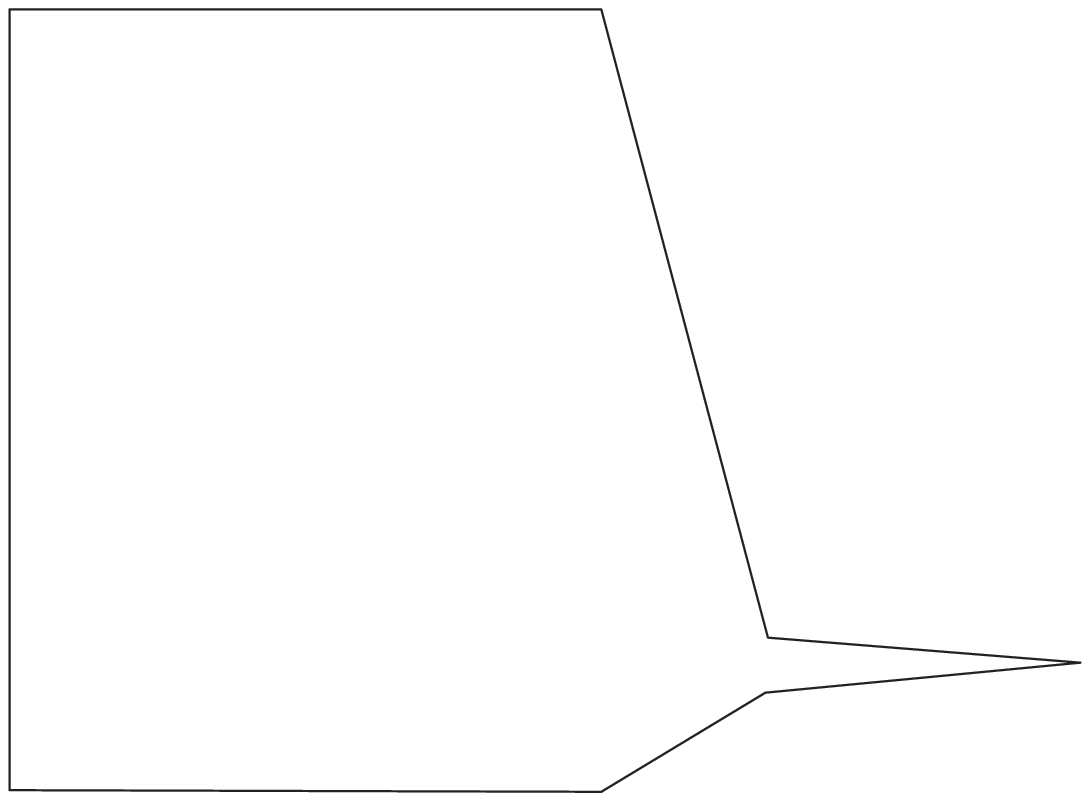}}
  \end{picture}
 \caption{Ge detector inside the active (NaI, PS) and passive (Cu, Pb) shielding.}
 \label{Fig.setup}
\end{figure}

The background is suppressed in several steps. First, the detector is placed inside a cup-like NaI crystal with 14 cm thick walls surrounded with 5 cm of electrolytic copper and 15 cm of lead (Fig.~\ref{Fig.setup}). Active and passive shielding reduces the external $\gamma$-background in the ROI to a level of $\sim2$~counts/keV/kg/day. Being located just under reactor \#2 of the KNPP (at a distance of 13.9 m from the reactor core, which corresponds to an antineutrino flux of $2.7\times 10^{13}\;{\rm \bar\nu_e/cm^2/s}$), the detector is well shielded against the hadronic component of cosmic rays by the reactor body and technological equipment (overburden$\simeq$70 m w.e.). The muon component is also reduced by a factor of $\sim$10 at $\pm20^\circ$ with respect to the vertical direction and of $\sim$3 at $70^\circ\!\ldots80^\circ$, but a part of the residual muons are captured in the massive shielding and thus produce neutrons which scatter elastically in Ge and give rise to a low-energy background. To suppress this background, the spectrometer is covered with additional plastic scintillator plates (PS) which produce relatively long $\mu$-veto signals. Special care is taken to reduce non-physical low-amplitude circuit noise (afterpulses, radio frequency interference, microphonism, etc.). Thus, for example, the detector signal is processed by three parallel independent electronic channels with different shaping time (2, 4 and 12 $\mu$s), which allows a primitive Fourier analysis \cite{Garcia92} to be performed \`{a} posteriori, so artefact signals are discriminated (see the next section).

\section{Data taking and
processing}

In order to get a recoil electron spectrum, we use a differential
method comparing the spectra measured during the reactor operation (ON)
and shutdown (OFF) periods. In our previous work we considered Phase-I (13 months'
measurement from 08.2005 to 09.2006, including 5184 and 1853
ho\-urs of the reactor ON and OFF periods, respectively). Today we can add Phase-II -- 19 months from 09.2006 to 05.2008.
Unfortunately, for some organizational and technical reasons, there were several interruptions in the measurement. After the preliminary selection described below, 6798 ON-hours and 1021 OFF-hours of active time were found to be available for analysis.

During the measurements, the signals of the HPGe detector,
anticompton NaI shielding and outer anticosmic plastic counters, as well as the dead-time information, are collected on an event-by-event basis.
The neutrino flux monitoring in the ON period is carried uot via the reactor
thermal power measured with an accuracy of 0.7\%.

The collected data are processed in several steps. The first step involves different selections aimed at suppressing nonphysical and physical backgrounds:
\begin{enumerate}
\item {\bf Bad run} rejection. We reject those one-hour runs which correspond to the periods of liquid nitrogen filling and any mechanical or electrical work at the detector site, as they could produce noise.
\item {\bf Radioactive noble gas} rejection. Unfortunately, the detector shielding turned out to be not tight enough against radioactive noble gases. To smooth away this design defect, we analyze energy spectra measured during each several hours and check the stability of the $\gamma$-background. If any visible excess of 81~keV ($^{133}$Xe), 250~keV ($^{135}$Xe) or 1294~keV ($^{41}$Ar) $\gamma$-line occurs, the corresponding runs are removed.\footnote{In fact, these files are used later for the ``noble gas'' correction of the rest of the data.}
\item {\bf Detector noise} rejection. For some obscure reasons our Ge detector happened to become noisy from time to time. In order to reject these noisy periods, the low-amplitude count rate is checked second by second, and those seconds which contain more than 5 events with $E>$2~keV are rejected.
\item {\bf Audio-frequency} rejection. We reject those events which are separated by a time interval shorter than 80~ms or equal to ($n\cdot20.0\pm0.1$)~ms. In such a way we suppress the noise caused by mechanical vibrations (``ringing'') and the 50~Hz power-line frequency.
\item {\bf Fourier} rejection. As it was mentioned, the real and artefact signals have different Fourier spectra. To exploit this difference, we build three plots similar to that shown in Fig.~\ref{Fig.Matrix}: ($E_2$ vs $E_1$), ($E_3$ vs $E_2$) and ($E_1$ vs $E_3$).

\begin{figure}[hb]
 \setlength{\unitlength}{1mm}
  \begin{picture}(140,65)(0,0)
   \put(0,-3){\includegraphics{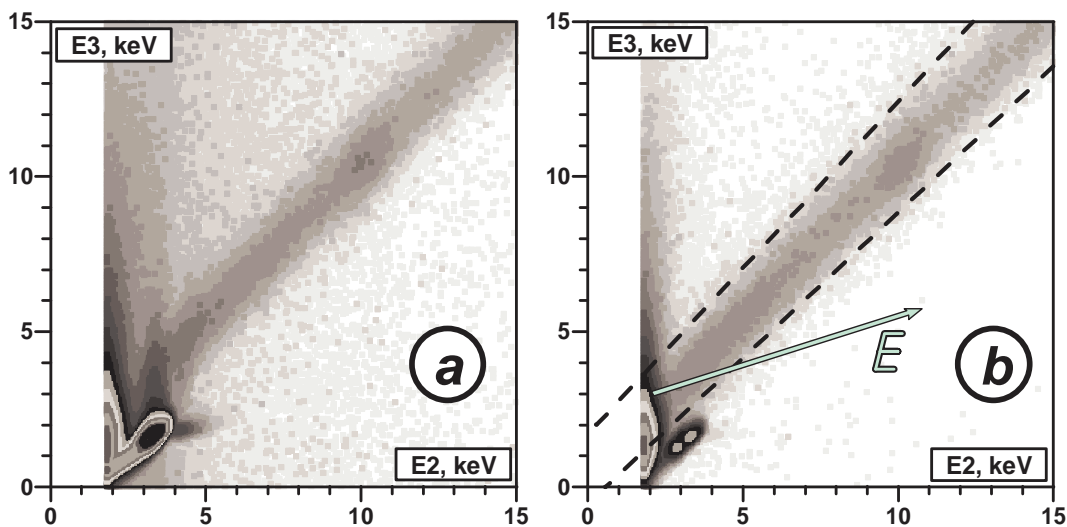}} 
  \end{picture}
 \caption{Example of the Fourier analysis made with different shaping-times: ADC-2 operates with 4~$\mu$s pulses, and ADC-3 operates with 12~$\mu$s pulses. Plot (a) is made {\em before} and (b) -- {\em after} the ``audio-frequency'' rejection; it is seen that most of the rejected events are non-diagonal. (The color intensity scale is logarithmic.)}
 \label{Fig.Matrix}
\end{figure}

    The real signal falls into diagonals ($E_1\simeq E_2\simeq E_3$) within the energy resolution, whereas any nonphysical artefact shows a different pattern. We only select {\em diagonal} events and thus additionally reject low- and high-frequency noise. To ensure the lowest cut-off, we replace $E_1$, $E_2$ and $E_3$ by their linear combination $\vec{E}$:
    \begin{equation}\label{Eq.E(abc)}
    \vec{E}=aE_1+bE_2+cE_3\;\;,
    \end{equation}
    where the amplitudes $a$, $b$, $c$ are chosen (subjectively) so as to make the vector $\vec{E}$ be antiparallel to the noise gradient (Fig.~\ref{Fig.Matrix}b).
\end{enumerate}

After the above rejections we construct energy spectra for the ON and OFF periods and correct\footnote{The corrections do not introduce a significant error in the final result as they affect both ON and OFF spectra in the same way.}
 them in two steps:
\begin{enumerate}
\item {\bf Noble gas} correction. As our spectrometer is not located in a special laboratory, but in the {\em technological room}, there are sometimes short operational periods when the concentration of $^{41}$Ar, $^{133}$Xe and $^{135}$Xe in this room becomes higher than usual. Spectra measured under these conditions are used to evaluate the contribution of each radioactive gas to the low-energy part of the background. These contributions normalized to the intensities of the corresponding $\gamma$-lines are then subtracted from those few ON and OFF spectra where small traces of these lines are still present. In this case the value of such correction in the ROI\footnote{The Region-Of-Interest (ROI) in our analysis includes two fragments from 3 to 9.4 and from 11.2 to 55 keV, i.e. the low-energy part of the continuous spectrum without peaks which could depend on the reactor operation.} does not exceed 1-2\%.
\item {\bf Low-energy threshold} correction. The detection efficiency $\eta$ just above the threshold $E_0$ is measured with a pulser and fitted with the function
    \begin{equation} \label{eq.threshold}
    \eta(E) = \int\limits_{-\infty}^E\frac{1}{\sqrt{2\pi}\,\sigma} e^{-\frac{\left(x-E_0\right)^2}{2\sigma^2}}dx
    \end{equation}
    where $\sigma$ stands for the detector energy resolution. Experimental spectra are then corrected by the function (\ref{eq.threshold}) which in our case becomes significant at energies below 2.8 keV.
\end{enumerate}

As a result, we obtain energy spectra $S$ for the ON and OFF periods which must be normalized to the corresponding active times $T_{\rm ON}$ and $T_{\rm OFF}$ and then compared to each other, taking into account the additional neutrino-dependent term:
\begin{equation}\label{eq.ON=OFF+nu}
\begin{array}{ccccc}
\underbrace{\hspace{3mm}\frac{S_{\rm ON}}{T_{\rm ON}}\hspace{3mm}}\hspace{3mm}&=&\underbrace{\hspace{3mm}\frac{S_{\rm OFF}}{T_{\rm OFF}}\hspace{3mm}}&+&\hspace{3mm}\underbrace{\hspace{3mm}m_d\;\Phi_\nu \left(W + X\cdot EM\right)\hspace{3mm}}\\
{\rm (effect)}&&{\rm (background)}&&{\rm (neutrino\;\;\; contribution)}
\end{array}
\end{equation}
This term includes the fiducial detector mass $m_d$ and the antineutrino flux $\Phi_\nu$ (known with an accuracy of 1.7\% and 3.5\%, respectively) multiplied by the sum of two neutrino contributions: the weak one ($W$), which can be calculated easily by formula (\ref{eq.dsW/dT}) and is completely negligible in our case, and the electromagnetic one ($EM$), which is proportional to the squared NMM value:
\begin{equation}\label{eq.X}
X\equiv \left(\frac{\mu_\nu}{10^{-11}\mu_{\rm B}}\right)^2 \;.
\end{equation}

Unfortunately, the ON and OFF periods are not equal from the point of view of statistics. A usual OFF period is much shorter, and, therefore, the final sensitivity is limited by the background uncertainties. On the other hand, today, after three years of data taking, we know the ROI background structure with more confidence. It gives us the right to introduce additional information in our analysis, namely, to state that our background is {\sl a smooth curve}.

To implement this conventional idea, we fit the background OFF spectrum in the ROI from 2.9 keV to 55 keV with a parametrized smooth function (a sum of Gaussian, exponential and linear functions) or with splines; these fits produce slightly different results, and their spread is taken into account by the final systematic error.

Then we compare the ON spectrum channel by channel with the obtained background curve and extract the $X$-value (or its upper limit) from Eq.(\ref{eq.ON=OFF+nu}). This evaluation is more complicated than expected because it is very difficult to count active times $T_{\rm ON}$ and $T_{\rm OFF}$ precisely in a proper way (especially, after numerous selections of the events). To avoid possible errors caused by this procedure, we divide the active time normalization into two parts: absolute ($T_{\rm ON}$) and relative
($\tau\equiv T_{\rm ON}/T_{\rm OFF}$).

Roughly both the $T_{\rm ON}$ and $T_{\rm OFF}$ active times are estimated using several background $\gamma$-lines: the 238~keV line of $^{212}$Pb, the 1173~keV and 1333~keV lines of $^{60}$Co and the 1461~keV line of $^{40}$K. This radiation originates from the pollution of the internal parts of the spectrometer and is therefore more-or-less stable in time and does not depend on the reactor operation. Comparing the intensities of the above lines measured {\em with} and {\em without} any selections, we get the estimates $\widetilde{T}_{\rm ON}$, $\widetilde{T}_{\rm OFF}$  with accuracy of 0.9\% and 1.9\%, respectively. This is not enough, however, to evaluate the $\tau$ value with the required precision. We resolve this problem in the following way.

The relative ON/OFF time factor $\tau$ is represented as a product of its estimate $\widetilde{\tau}=\widetilde{T}_{\rm ON}/\widetilde{T}_{\rm OFF}$ (which is a constant known with a bad accuracy of 2.1\%) and a correction factor $K$ (which should be not far from 1.0):  $\tau=K\widetilde{\tau}$. Then Eq.~(\ref{eq.ON=OFF+nu}) can be transformed to
\begin{equation}\label{eq.nu=ON-kOFF}
 X\cdot EM+W=\left(S_{\rm ON}-K\cdot\widetilde{\tau}\cdot S_{\rm OFF}\right)\cdot\left(T_{\rm ON}\;m_d\;\Phi_\nu\right)^{-1}
\end{equation}
As it is seen, the absolute time normalization $T_{\rm ON}$ contributes to the final result in the same way as $\Phi_\nu$ or $m_d$ (i.e., simply as a {\em factor}) and therefore we can replace $T_{\rm ON}$ by its estimate $\widetilde{T}_{\rm ON}$. Standard systematic deviation $\delta X$ caused by this factor is not significant:
\begin{equation}\label{Eq.d(Fi,m,T)}
\frac{\delta X}{X}=\frac{\delta(T_{\rm ON}\cdot m_d\cdot\Phi_\nu)}{T_{\rm ON}\cdot m_d\cdot\Phi_\nu}=\sqrt{(0.9\%)^2+(1.7\%)^2+(3.5\%)^2}\simeq4.0\%
\end{equation}

\begin{figure}[ht]
 \setlength{\unitlength}{1mm}
  \begin{picture}(140,65)(0,0)
   \put(6,5){\includegraphics{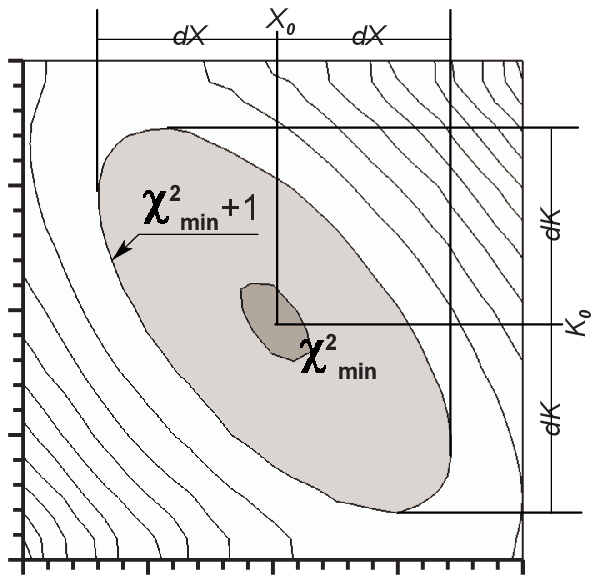}}
   \put(2,55){\mbox{\footnotesize\sf K}}
   \put(0,44.0){\mbox{\scriptsize\sf 1.01}}
   \put(0,31.5){\mbox{\scriptsize\sf 1.00}}
   \put(0,19.0){\mbox{\scriptsize\sf 0.99}}
   \put(0, 6.5){\mbox{\scriptsize\sf 0.98}}
   \put(57.5,2.5){\mbox{\footnotesize\sf X}}
   \put(4.5,3){\mbox{\scriptsize\sf --100}}
   \put(18.0,3){\mbox{\scriptsize\sf --50}}
   \put(32.5,3){\mbox{\scriptsize\sf 0}}
   \put(44.5,3){\mbox{\scriptsize\sf 50}}
   \put(80,5){\includegraphics{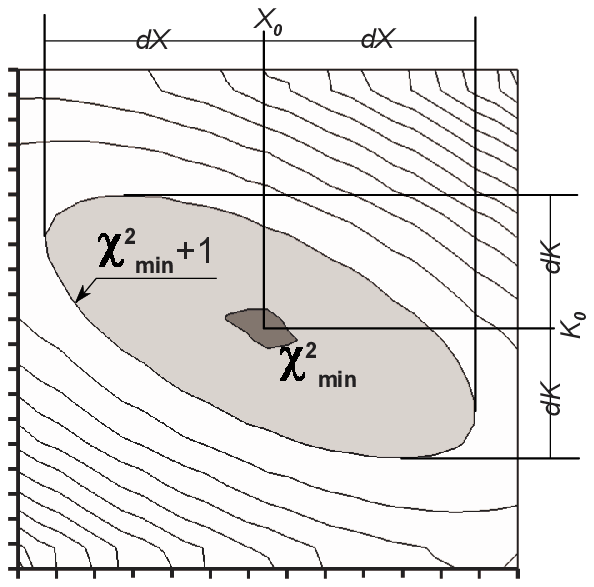}
    \put(-64,50){\mbox{\footnotesize\sf K}}
    \put(-66,38.0){\mbox{\scriptsize\sf 1.01}}
    \put(-66,25.5){\mbox{\scriptsize\sf 1.00}}
    \put(-66,13.0){\mbox{\scriptsize\sf 0.99}}
    \put(-66, 0.5){\mbox{\scriptsize\sf 0.98}}
    \put(-8.5,-2.5){\mbox{\footnotesize\sf X}}
    \put(-60.5,-2){\mbox{\scriptsize\sf --80}}
    \put(-46.0,-2){\mbox{\scriptsize\sf --40}}
    \put(-27.5,-2){\mbox{\scriptsize\sf 0}}
    \put(-13.5,-2){\mbox{\scriptsize\sf 40}}
   } 
  \end{picture}
 \caption{Example of two-dimensional charts $\chi^2(X,K)$ built for some portions of the GEMMA data.}
 \label{Fig.Chi^2(X,K)}
\end{figure}

Considering $X$ and $K$ in Eq.(\ref{eq.nu=ON-kOFF}) as two unknown free parameters and varying them around the point ($X$=0; $K$=1), we calculate residual sums $\chi^2$ and thus build a two-dimensional chart $\chi^2(X,K)$ similar to the one shown in Fig.~\ref{Fig.Chi^2(X,K)}. The appearance of a closed graph in the chart corresponds to the minimum of $\chi^2$. the non-zero eccentricity of the ellipse-like curve indicates some non-zero correlation between $X$ and $K$, but this correlation is not too high, so that both $X$ and $K$ must be fitted only together, whereas can be preliminarily estimated separately.

\begin{figure}[ht]
 \setlength{\unitlength}{1mm}
  \begin{picture}(140,50)(0,0)
   \put(0,-3){\includegraphics{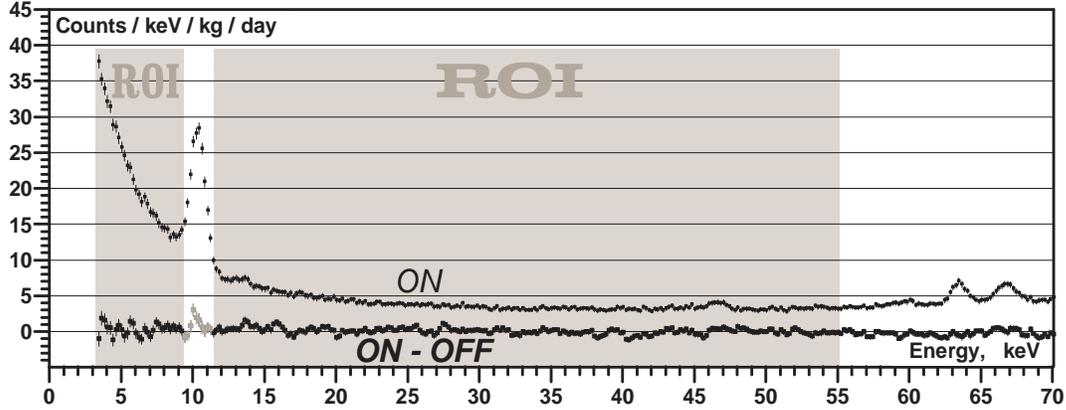}} 
  \end{picture}
 \caption{Fragments of the experimental ON and ON--OFF spectra measured in Phase-II (only a part of the available statistics is presented).}
 \label{Fig.ON-spk}
\end{figure}

Indeed, according to Fig.~\ref{Fig.W/EM}, only the left-hand part of the ROI is really sensitive to the $X$-value, whereas the right-hand part is almost insensitive to $X$ but still sensitive to the relative time factor $\tau$ and hence to the $K$ parameter. It means that preliminary time normalization can be performed not only with the background $\gamma$-lines, but also with a part of the continuous spectrum (e.g., from 20 to 55 keV). Both methods give very similar results, but the second one provides better precision. The differential ON--OFF spectrum built with this $\tau$ estimate is shown in Fig.~\ref{Fig.ON-spk}.

The gain, energy scale, resolution and noise level were not exactly the same for different long-term ON measurement periods of Phase-II. As a result, the optimal coefficients $a$, $b$ and $c$ of Eq.~(\ref{Eq.E(abc)}), as well as the low-energy edge of the ROI, were also changed. The data of these periods must be analyzed separately and only then their results be combined in the final distribution. The analysis of the $j$-period consists in channel-by-channel comparison of the ON and OFF spectra over the ROI (with allowance for the weak contribution) and building a $\chi^2(X,K)$ chart similar to the one shown in Fig.~\ref{Fig.Chi^2(X,K)}. Projection (not a section!) of the whole chart onto the $X$-axis gives us the $X_j$ value with a standard statistical deviation $dX_j$. Taking into account the standard systematic deviation $\delta X_j$ given by Eq.(\ref{Eq.d(Fi,m,T)}), we get $X_j\pm dX_j\pm\delta X_j$ and build the probability curve $p_j(X)$ which already includes the $K$-uncertainty. Multiplying probability curves of all analyzed periods (and also of both Phases I and II), we get the probability distribution $P(X)=\prod p_j(X)$  for the total measurement.

\begin{figure}[ht]
 \setlength{\unitlength}{1mm}
  \begin{picture}(140,40)(0,0)
   \put(-0.5,-3){\includegraphics{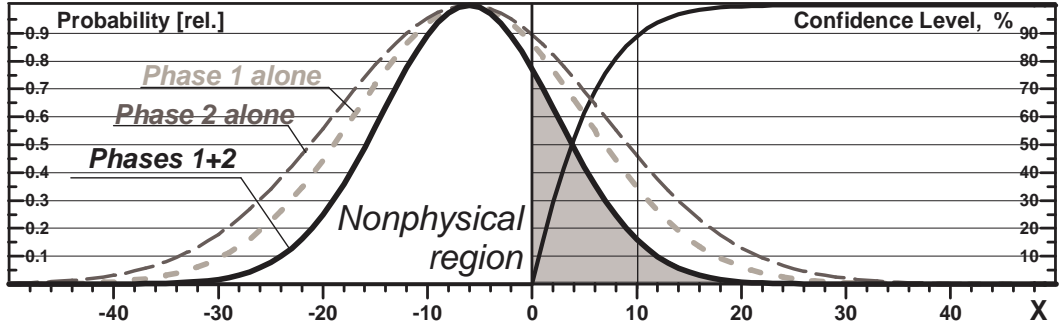}} 
  \end{picture}
 \caption{Extraction of the NMM limit with the required CL from the renormalized part of the probability distribution $P(X)$.}
 \label{Fig.pro_FE_b}
\end{figure}

In order to estimate the part of systematic error introduced by the subjective choice of the background approximation function, ($a$,$b$,$c$)-coefficients and ROI edges, we repeat the above procedure several times for different reasonable combinations of these parameters and build {\em a sum} of the obtained distributions (Fig.~\ref{Fig.pro_FE_b}) which turns out to be about 7\% wider. After a conventional renormalization recommended by the Particle Data Group\cite{PDG} and described in our previous work\cite{Beda04}, we extract the upper limit for the $X$ parameter and thus get the following NMM limit:
\begin{equation} \label{Eq.3.2}
\mu_\nu<3.2\times10^{-11}\mu_{\rm B}\hspace{5mm}{\rm (90\%CL)}
\end{equation}

\section{Atomic ionization effect}
\begin{figure}[th]
 \setlength{\unitlength}{1mm}
  \begin{picture}(140,60)(0,0)
   \put(0,5){\includegraphics{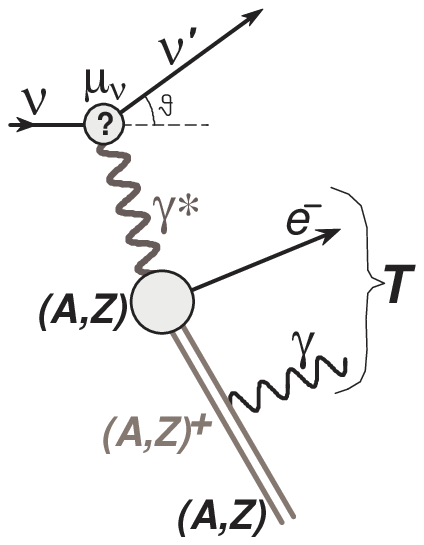}}
   \put(53,0){\includegraphics{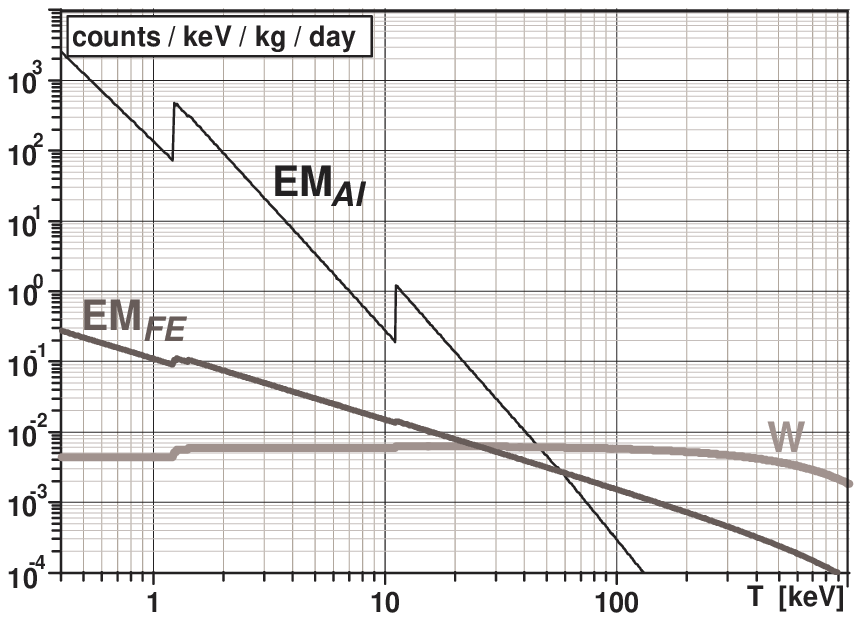}}
  \end{picture}
 \caption{The NMM interaction via the atomic ionization (AI) channel. Left: scheme of the process. Right: expected GEMMA count rates calculated for $\mu_\nu=3.2\times10^{-11}\mu_{\rm B}$ and compared to the weak (W) and electromagnetic ($\nu,e$) scattering on free electrons (FE).}
 \label{Fig.Idea_AI}
\end{figure}

When this work was almost ready, H.N.~Wong et al.\cite{Wong10} noticed one more channel of the NMM interaction, namely, atomic ionization (AI)
\begin{equation}
\begin{array}{ccccccl}
 \nu + (A,Z)&\longrightarrow&\nu\;'     &+&(A,Z)^+    &+&e^-\\
            &           &           & &\multicolumn{3}{l}{\rule{4mm}{0mm}\downarrow \rm recombination}   \\
            &           &           & &\!\!\!(A,Z)      &+&\gamma
\end{array}
\end{equation}
The AI channel (Fig.~\ref{Fig.Idea_AI}) dominates at very low transferred energies $T$ when the wave length of the virtual photon $\lambda(\gamma*)\!=\!hc/T$ is comparable with the atomic dimension.
In this case (i.e., at $T\sim1$ keV), the virtual photon may interact not only with an atomic electron considered free, but also with the atom as a whole. According to \cite{Wong10}, the cross section of this process could be described as
\begin{equation}
{d\sigma_{AI}}/{dT} \simeq\frac{r_0^2}{\pi\alpha} \left( {\frac{\mu _\nu }{\mu _{\rm B} }}
\right)^2 \frac{E_\nu^2}{T}\; \sigma_{\gamma A}\;\;,
\label{Eq.AI}
\end{equation}
where $\sigma_{\gamma A}$ stands for the photoelectric cross section of a physical photon with the energy $E_\gamma=T$.

The right-hand part of Fig.~\ref{Fig.Idea_AI} shows the GEMMA count rates caused by weak (W) and electromagnetic (EM$_{AI}$ and EM$_{FE}$) interactions calculated for  $\mu_\nu=3.2\times10^{-11}\mu_{\rm B}$. As one can see, the EM$_{AI}$ is two to three orders of magnitude higher\footnote{It should be mentioned that according to another approach\cite{Voloshin10}, the AI effect is not so high and even may be negligible.} than the EM$_{FE}$ in the low-energy region. It means that from the point of view of the NMM-sensitivity, the region ($T\leq$10 keV) is especially important, at least while the wave length  $\lambda(\gamma*)$ is rather short with respect to the crystal structure.

\begin{figure}[ht]
 \setlength{\unitlength}{1mm}
  \begin{picture}(140,50)(0,0)
   \put(0,-3){\includegraphics{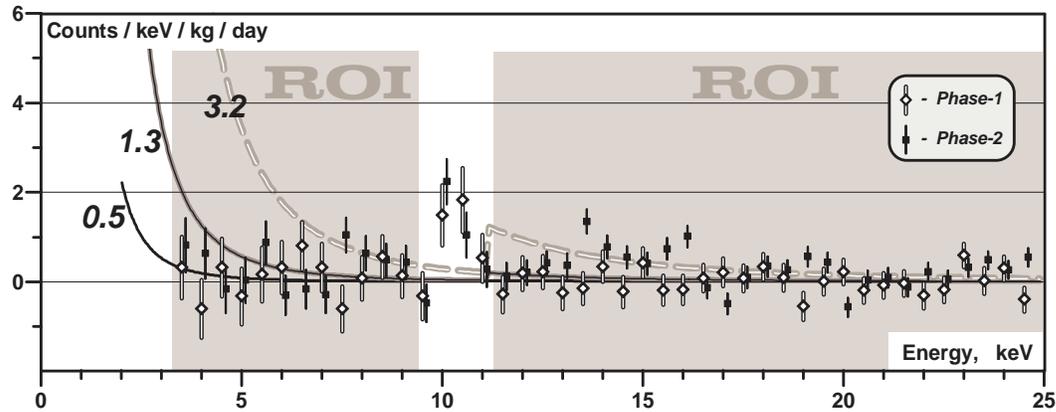}} 
  \end{picture}
 \caption{Differential ON--OFF spectra (points) measured in Phases I, II and expected count rates (curves) calculated with formula (\ref{Eq.AI}) for different NMM values in terms of 10$^{-11}\mu_{\rm B}$.}
 \label{Fig.AI}
\end{figure}

Unfortunately, our GEMMA spectrometer was not adjusted for the detection of $T\!\leq3$ keV. Nevertheless, we estimated its sensitivity to the AI process. Figure~\ref{Fig.AI} represents our experimental points measured in Phases I and II compared to the theoretical curves calculated with formula (\ref{Eq.AI}) for different NMM values.

\begin{figure}[hb]
 \setlength{\unitlength}{1mm}
  \begin{picture}(140,50)(0,0)
   \put(0,-3){\includegraphics{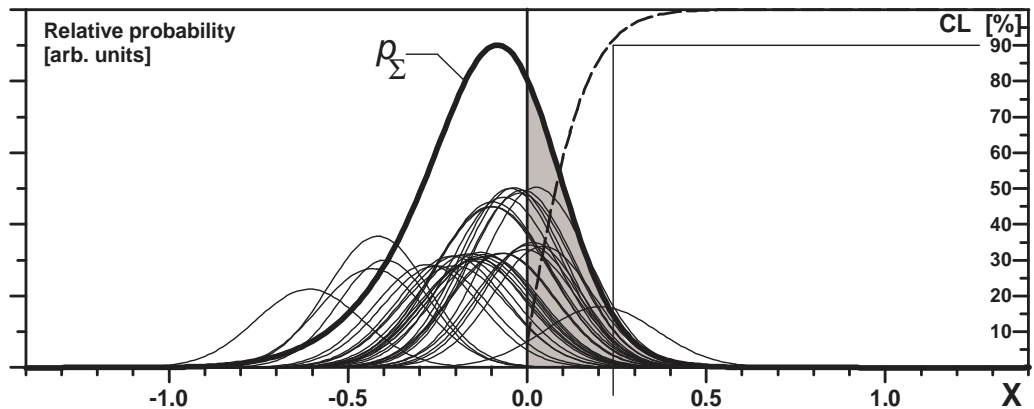}} 
  \end{picture}
 \caption{Probability distributions obtained for different subjective parameters used in the analysis; $P_\Sigma$ represents their weighted sum.}
 \label{Fig.AI_probability}
\end{figure}

Exact mathematical analysis of both Phase I and Phase II results shows that the statistical precision of our data allows reaching $(2\ldots3)\times10^{-12}\mu_{\rm B}$ (at the 90\%CL). At the same time, the possible systematic error (and, therefore, the wide spread of the separate curves of Fig.~\ref{Fig.AI_probability}) is much higher. It originates mainly from the fact that the "hard" energy threshold of the detection system in Phases I and II was not low and stable enough. Under this condition, it is only possible to set the following NMM limit:
\begin{equation} \label{Eq.0.5}
\mu_\nu<5.0\times10^{-12}\mu_{\rm B}\hspace{5mm}{\rm (90\%CL)}
\end{equation}

\section{Conclusion}

The experimental NMM search with the GEMMA spectrometer has been going on at the
Kalinin Nuclear Power Plant (Russia) since 2005. The HPGe detector of 1.5 kg placed  13.9~m under the core of the 3~GW$_{\rm th}$ water-moderated reactor is exposed to the antineutrino flux of $2.7\!\times\!10^{13}\;\bar\nu_e/{\rm cm}^2/{\rm s}$. As a result of the 3-year measurement (about 13000 ON-hours and 3000 OFF-hours of active time), the upper limits of 5.0$\times$10$^{-12}\mu_{\rm B}$ and 3.2$\times$10$^{-11}\mu_{\rm B}$ at 90\% CL were found for the NMM with and without using the AI-interaction mechanism, respectively. As it was mentioned by M.B.~Voloshin \cite{Voloshin10} contrary to H.T.~Wong et al. \cite{Wong10}, the atomic contribution is not essential at all, so that the first of the above limits should be considered as illustrative only.

At present, analysis of the data taken under improved conditions (Phase III) has just started, but it indicates that further sensitivity improvement of the spectrometer can be reached only by its significant upgrading. Within the framework of this new project (GEMMA-2) we will use the antineutrino flux of $\sim5.4\times10^{13}\; \bar\nu_e /{\rm cm}^2/{\rm s}$, increase the
mass of the germanium detector by a factor of four and decrease the level of the background.
The main improvement is expected to be the significant lowering of the energy threshold (below 1 keV).
These measures will provide the possibility of achieving the NMM limit at the level of $1.0\times10^{-12}\mu_{\rm B}$.

\begin{ack}
The authors are grateful to M.B.~Voloshin for his important comments and to administration of the KNPP and the staff
of the KNPP Radiation Safety Department for permanent assistance in the
experiment.
This work is supported by the Russian State Corporation ROSATOM
and by the Russian Foundation for Basic Research, pro\-jects 09-02-00449 and 09-02-12363.
\end{ack}

\end{document}